Bogdan Czejdo , Sambit Bhattacharya, Mikołaj Baszun, Wiktor B. Daszczuk

# IMPROVING RESILIENCE OF AUTONOMOUS MOVING PLATFORMS BY REAL-TIME ANALYSIS OF THEIR COOPERATION

***Abstract.*** *Environmental changes, failures, collisions or even terrorist attacks can cause serious malfunctions of the delivery systems. We have presented a novel approach improving resilience of Autonomous Moving Platforms AMPs. The approach is based on multi-level state diagrams describing environmental trigger specifications, movement actions and synchronization primitives. The upper level diagrams allowed us to model advanced interactions between autonomous AMPs and detect irregularities such as deadlocks live-locks etc. The techniques were presented to verify and analyze combined AMPs' behaviors using model checking technique. The described system, Dedan verifier, is still under development. In the near future, a graphical form of verified system representation is planned.*

## INTRODUCTION

The growing scope of applications of new autonomous mobile devices must include the development of the resilient technologies to respond properly to environmental changes, failures, collisions or even terrorist attacks. The development of resilient systems of cooperating autonomous mobile platforms (AMPs) should have high priority since they are being applied in many areas such as: personal rapid transit (PRT)[1][2][3], marine seaport transportation systems[4], military automated transportation systems, military reconnaissance, surveillance and guard systems, and many others.

For all of these areas, models can be built to significantly increase the resilience of the involved systems. More specifically, for each application area we considered the following research problems:

1. Individual Autonomous Moving Platforms and fail-safe techniques for them. Within this problem we considered development of failure detection and failure avoidance techniques for individual AMPs.
2. Resilient cooperation of Autonomous Mobile Platforms. Within this research problem we considered building enhanced cooperation algorithms. That mainly included developing techniques for cooperation, failure detection, and cooperation modification.
3. Resilient cooperation of Autonomous Mobile Platforms with changes in an Environment. This research problem required building ontology for environment and its changes, and development of techniques for change detection and cooperation modification.

The research was integrated vertically and horizontally. The horizontal integration of research meant that the similar research problems for various application areas were analyzed and solutions from one application area assisted work in another area. Similarly the vertical integration of research was pursued to make sure that solution for individual problems provide the proper solutions to the whole application area.

The result of horizontal integration and vertical integration lead us to identify common research techniques. These common research techniques can be described by the following research questions:

A. How can ontology for the needed resources be designed so that the environment changes can be reflected?
B. How can deterministic state diagrams be used across various AMPs and various cooperation algorithms for AMPs?
C. How critical is simulation for various algorithms of cooperation of autonomous moving platforms?
D. How feasible is real-time verification of correctness of cooperation of autonomous moving platforms?

In this paper we concentrate on the problems of Autonomous Mobile Platforms navigation in an outdoor environment [5][6]. We assume that the AMP not only responds directly to the environment [7][8] but also to actions of other AMPs [9]. State diagrams [10][11] have been previously used to describe the AMP behavior [12]. Typically, the appropriate software is developed manually based on such models. To accelerate the development process we have created a new tool for the design of AMP behavior and verification (Dedan). Such a tool can be very useful for the rapid modification of AMP reactive behavior. We are working on extending the tool functionality to automatically generated AMP behavior in response to changing requirements [13].

In this paper we describe the techniques to analyze state diagrams, and for integration of multiple state diagrams. The state diagrams allowed us to model advanced interactions between autonomous AMPs and can ensure the correctness of AMP interactions. When rapid modifications of AMP behavior are required, the rapid checking of AMP interactions is crucial. The model checking method for state diagrams can identify problems such as deadlocks or live-locks and therefore offer the AMP designer a set of ready-to-use algorithms and techniques for the rapid system behavior modification.

The organization of the paper is as follows. In Section 1 we describe a sample outdoor environment referred to as environmental resources. The state diagrams and environmental triggers are described in Section 2. In Section 3, we describe state diagrams for cooperating AMPs. In Section 4, we show how state diagrams can be analyzed and modified for increased resilience.

## 1. ENVIRONMENTAL RESOURCES

From the point of view of the AMP programmer it is important to know the type of an environment the AMPs will move through. In general, an environment can be known or unknown. In this paper we will concentrate on describing AMP behavior in a known environment. The known environment is typically described by a geographic map identifying all landmarks and roads with the characteristics anticipated by the AMPs movement. One of the simple representations of a topological map can be a graph showing all accessible places in the form of nodes designating road (and parking) markers and the ways to get to these places in the form of graph paths designating existing roads. Each road fragment between road markers can have a one lane, two lanes, or multi-lane characteristic.

Any topological map in the form of a graph can be also interpreted as a graph of environmental resources. It means that each node of the graph can be also interpreted as a resource and when the AMP position is associated with this node we can claim that the AMP acquired the resource. When the AMP leaves the node we say that it releases the resource. The link between two nodes can be also interpreted as resource that can be acquired and released. Such an interpretation of a topological graph allows us to apply known resource allocation algorithms for the description of multiple AMP behavior.

Let us assume, for our case study, the outdoor environment includes a road with road fragments defined by three road markers leading to three warehouse lots. The corresponding environmental resource graph can be constructed as shown in Fig. 1.

The following nodes corresponding to environmental resources can be identified, using letter E for edge resources and M for middle resources: (R1) "Warehouse Lot E1", (R2) "Road Marker E1", (R3) "Road Marker M", (R4) "Road Marker E2", (R5) "Warehouse Lot E2", and (R6) "Warehouse Lot M". The road fragments can be also designated explicitly as resources but in this paper we designated them implicitly.

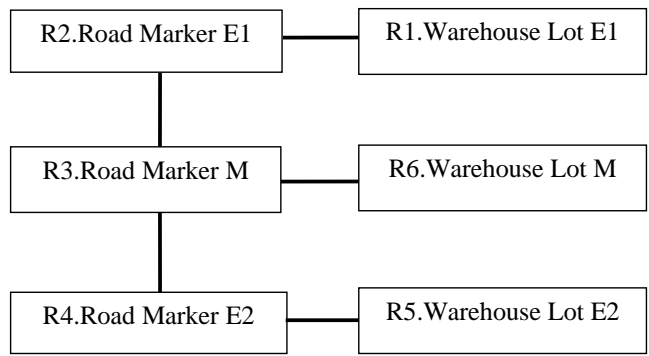

*Fig. 1 An Environmental Resource Graph with 6 resources*

## 2. STATE DIAGRAMS WITH TRANSITIONS FOR AMP NAVIGATION

The deterministic state diagrams are well described in literature [10][11]. Generally, the deterministic state diagram, in addition to states, has transitions consisting of triggers that cause the transition of the AMP from one state to another, and actions, that are invoked during a transition. Triggers are expressed by Boolean conditions evaluated continuously to respond to changes in the environment.

To specify state diagrams we use the notation based on Universal Modeling Language (UML) [14] where a state is indicated by a box and a transition is indicated by an arrow with a label. The first part of the transition label (before the slash) specifies the trigger and the part after the slash specifies the action (or message) to be invoked during the transition [14]. The syntax of probabilistic specifications is described in the literature [15] as an additional third component specifying the probability of the entire transition.

Deterministic state diagrams are well described in literature [10][11]. Generally, the deterministic state diagram, in addition to states, has transitions consisting of triggers that cause the transition of the AMP from one state to another, and actions, that are invoked during a transition. Triggers are expressed by Boolean conditions evaluated continuously to respond to changes in the environment.

To specify state diagrams we use the notation based on Universal Modeling Language (UML) [14] where a state is indicated by a box and a transition is indicated by an arrow with a label. The first part of the transition label (before the slash) specifies the trigger and the part after the slash specifies the action (or message) to be invoked during the transition [14].

State diagrams that are explicitly location dependent can be convenient to specify AMP behavior for several reasons. Firstly, the diagram can be constructed by relatively simple transformation of environmental resource diagram. Second, probabilistic components can be added relatively easily. Thirdly, the behavior of cooperating AMPs can be described by concurrent state diagrams and all well-established techniques for concurrent program analysis can be used i.e. deadlock detection or deadlock avoidance algorithms. The analysis of concurrency can be done automatically and the AMP program can be directly generated from state diagram model.

Based on environmental graph and corresponding environmental triggers we can rapidly specify various location dependent state diagrams. For example, let us consider Behavior A describing a simple path for movement of AMP1: start from the Warehouse Lot E1, then follow the exit leading to the Road Marker E1, then continue to follow the road until encountering the road marker M, then still continue to follow the road until encountering the road marker E2 then enter the Warehouse Lot E2 and stop.

In order to model such behavior a state diagram model can be used. In general the multi-level model can be used but in this paper for the simplicity of presentation, we assume two level model. The upper level model is obtained by transforming the environmental graph i.e. converting non-directional to directional edges and providing the necessary triggers, actions and messages.

More precisely the link between two nodes e.g. "Road Marker E1" and "Road Marker M" can be interpreted as follows: if the AMP is assigned a resource "Road Marker E1" it should first acquire resource "Road Marker M" before releasing resource "Road Marker 1" account.

The state diagram shown in Fig. 2 specifies the AMP1 Behavior A in some detail dividing it into sequence of phases:

- Phase 1. Initially the AMP1 is in the "Warehouse Lot E1" state. In this state, if the needed resource is available i.e. road to Marker E1, then the transition takes place to the state "Road Marker E1". When the Road marker is reached then the state "At Road Marker E1" is also recorded as reached.
- Phase 2. In this state, if the needed resource is available i.e. road towards Marker M, then the transition takes place to the state "Moving towards Road Marker M". When the Road Marker M is recognized then the state "At Road Marker M" is reached.
- Phase 3. If possible then AMP1 transitions into "At Road Marker E2" to reach "At Road Marker E2" state and then finally
- Phase 4. It will get AMP1 into "At Warehouse Lot E2" state.

Similarly the AMP2 Behavior can be described into several phases allowing AMP2 to travel from Warehouse Lot E2 to Warehouse Lot E1.

In order to formally specify such phrases as shown in Fig. 2, we need topological identification triggers, topological actions, and synchronization messages. Let us describe them in this order. Each of these topological constructs can be defined by a lower level diagrams.

Different topological places i.e. different resources would usually generate different values for the AMP's sensors. The sensor signal processing algorithms i.e. algorithms describing a translation of AMP sensor signals into a high level signals that can be used to directly identify the environment. We will assume that a lower level state diagram can describe such algorithm and we will refer to these signals to be used by a higher level diagrams as the environmental triggers.

In our previous papers [13], we extensively studied the environmental triggers and their application to control the robot movements. Various computer vision techniques can be used to create such environmental triggers. The Hough transform is widely used in computer vision for detecting line segments and regular geometric features such as line segments and circles in images. More specifically, Progressive Probabilistic Hough Transform [16] can be used for detecting naturally occurring lines in images of roads[13]. The histogram based difference methods can be used for discriminating between road features and for recognition of major landmark objects [6].

Using Hough transform we can implement the environmental triggers to allow the AMP to direct itself to stay within the assigned lane. Histogram based difference measurement can be used during AMP navigation to solve the significant problem of recognition of landmark objects related to the road markers [6]. When the AMP moves beside a landmark the appropriate environmental trigger can generate *True* otherwise this condition is *False*. In general, in this paper we will assume an algorithm for self-driving vehicle with an ability to move the vehicle along sequence of specified road markers and to signal their recognized position.

To identify properly the solutions to our problems we will assume for further discussion the high level environmental identification trigger *detected(Road Marker X)*. This trigger reflects the ability of AMP sensors and algorithms to recognize the landmarks. The assumed high level environmental action corresponds to a movement of vehicle from the actual location to the provided Road Marker X e.g., *moveTo(Road Marker X)*. This action can be shown by a lower level diagram and using an environmental position marker *detected()* to determine the termination of the action.

Similarly, we assume the following environment synchronization triggers: *acquire(RX)* and *release(RX)*. The *acquire(RX)* trigger can be also defined by lower level state diagram with an environmental message *request_to_acquire*, then *wait* for available resource. The *release(RX)* is an environmental synchronization trigger generated to inform that the resource is not used any more.

## 3. INTERACTIONS OF AUTONOMOUS MOVING PLATFORMS

The state diagrams described in the previous section can be used to generate code for several interacting AMPs. In this section we discuss the use of state diagrams to describe and ensure the proper interactions between autonomous AMPs. Since in our approach we create explicit state diagrams based on environmental resource graph, we can take advantage of many theoretical and practical solutions in:

(a) geometrical modeling of AMP movement [5][6][7],
(b) resource allocation algorithms,
(c) model checking [16][17][18],

and apply them for verification of AMP behavior. Most solutions can be applied for both static verification of AMP behavior and the dynamic verification when the AMP is in the middle of execution of a program. The modeling of the AMP movement is typically related with spatial path description and can be done for both static and dynamic analyses. Resource allocation algorithms for deadlock avoidance can be used for dynamic verification to avoid collisions of moving AMPs assuming that they can wait in the state "*Moving towards…*" while doing their tasks.

The model checking provides a most general methodology [18][19][20][21][22] that can be used not only for deadlock avoidance or detection but also for detection and verification of wide variety AMP interaction characteristics. Typically the model checking is based on finite-state methods [20] that can be applied directly to our state diagrams and therefore it can be of important practical use for verifying AMP behaviors. The model checking method can offer the AMP behavior designer a set of ready-to-use algorithms and techniques for the analysis of complete system properties.

Conceptually, the approach is as follows. First, out of the specifications of the AMPs' behavior by state diagrams we build a possibly large but finite graph containing all possible (reachable) system states and all possible transitions among them. This graph defines the behavioral model of the set of AMPs. Each path in the graph represents an allowable execution or a (part of a) behavior of a system. The graph contains all possible executions or behaviors. The property list will be used for the graph correctness specification.

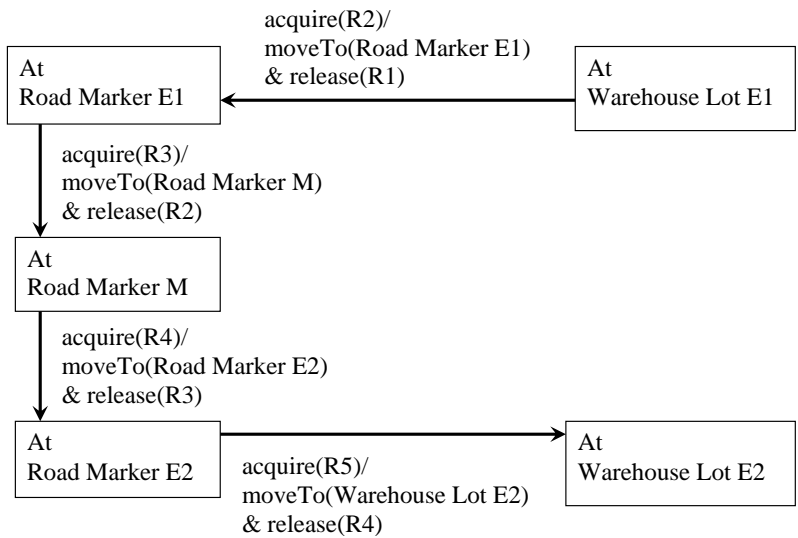

***Fig. 2.*** *State Diagram to describe Behavior of AMP1*

We have to deal with the exponential explosion of the state space size similar as others [18][21][22].There are many proposed and implemented solutions but the exponential growth of state space is still a real threat [17]. We included in our research a study of multiple forms of reduction of state space, aimed at removing the states and transitions which are irrelevant for the evaluation of a given formula. We also investigated the usefulness of compositional model checking, where some individual parts of a system (of a more acceptable size) are subject to an exhaustive state space search while the conclusion as to the performance of the whole system is reached by combining the results obtained for the individual parts.

Let us consider again the state diagram for AMP1 behavior as specified in Fig. 2. Let us assume that we have two AMPs. AMP1 behavior is exactly as in Fig. 2. AMP2 behavior is almost identical except that it starts from the Warehouse Lot E2 and terminates in Warehouse Lot E1. AMPs in general they can interact as in our example. The analysis of combined diagram might be necessary. There is a need for analytical transformations to investigate how the probability of reaching a given state by the first AMP affects the reachability of the states by the second AMP.

## 4. VERIFICATION IN DEDAN

For verification of the system presented in Figs. 1,2 and 3 it could be described in the Integrated Model of Distributed Systems (IMDS) formalism [23]. In this formalism a real distribution of elements may be expressed, since the actions of distributed elements are based on local states only. The Dedan verification environment, which uses IMDS specification, has been implemented to find deadlocks in cooperating distributed elements using model checking technique [24], Resource deadlocks and communication deadlock are searched automatically in Dedan.

A distributed system is typically described in terms of servers exchanging messages. A process in such a system can be defined as a sequence of changes of a server states. The states of servers are internal to the processes, which communicate by the message exchanges (a client-server model [25]).

Formally, a server state is a pair *p=(server, state)* and a message is a triple *m=(agent, server, service)*. An agent is an identifier distinguishing a distributed computation from other computations (can we talk about processes later). An action is a relation λ between an input pair *(m, p)* and an output pair *(m', p')*. A process is a sequence of actions: in the same server (server process) or in the same agent (agent process).

In the described system, we identify servers with static elements: the places of the AMPs environment (warehouse lots and road markers). Agents are identified with dynamic elements: AMPs travelling through the environment. For example, if the agent AMP1 is in the place Road Marker E1, then it tries to take the Road Marker M. To do it safely, first M is taken, and then E1 is released. It is done by means of three messages:

1. The message 'try' is sent from the E1 to M. This message may wait for acceptance for undefined period of time if M is occupied.
2. If at last the message 'try' is accepted in M is accepted (M is free at this time), the message 'ok' is sent back from M to E1. M changes its state from 'free' to 'reserved' – it cannot be taken by other AMP.
3. Then, E1 is released and M is finally taken by the AMP, E1 becomes 'free' and M becomes 'occupied'.

Processes of a concurrent system may fall into a deadlock. Dedan finds deadlocks in both views automatically and Presents them in readable form. Moreover, it finds partial deadlocks, in which not all of the servers are involved.

Yet, another model is possible: if a process is associated with an agent (rather than with a server), it migrates between servers and performs calculations in travelling way. It communicates with other travelling processes by means of servers' states. Messages are internal to a process. In such a way, a system is described in terms of resource sharing instead of message passing. This is similar to a Remote Procedure Calling (RPC) model [25] (yet, the analogy may be misleading because it is not necessary for a process to return to the calling server after an execution of a service).

The crucial fact is that it is the same system, shown in one of the two views, depending on connecting actions to form processes. If a sequence of actions is connected by server states - it is a server view. Server states are the carrier of the process, while messages are communication means. If a sequence is connected by messages – it is an agent view.

The deadlock may be observed from the agents' point of view. Dedan finds both kinds of deadlock (in communication and over resources) automatically.

In our example, a possibility of deadlock is obvious if two APMs travel, one from Lot E1 to Lot E3 and the other one travels opposite way. In the system, this „collision" of two AMPs can be solved in various ways. In general the solution can be based on:

- additional maneuvers maintaining the same route for all vehicles,

- changing the traveling route.

For our case study, the first possibility can be implemented using the Warehouse Lot M as a temporary parking space, where one of AMPs may wait for the other one passing M. This can be done by using additional message 'not' sent by E1/E2 if it is occupied by an AMP and the other AMP tries to take it standing at M. In such a situation, the message 'not' causes the AMP standing at M to divert to the Warehouse Lot M and let the AMP standing at E1/E2 to take M and drive on to E2/E1. Then, the AMP that stepped out of the way, drives from Warehouse Lot M back to M and continues its way. The modified diagram can be generated automatically as shown in Fig. 3.

The two-AMPs system is coded in IMDS as follows:

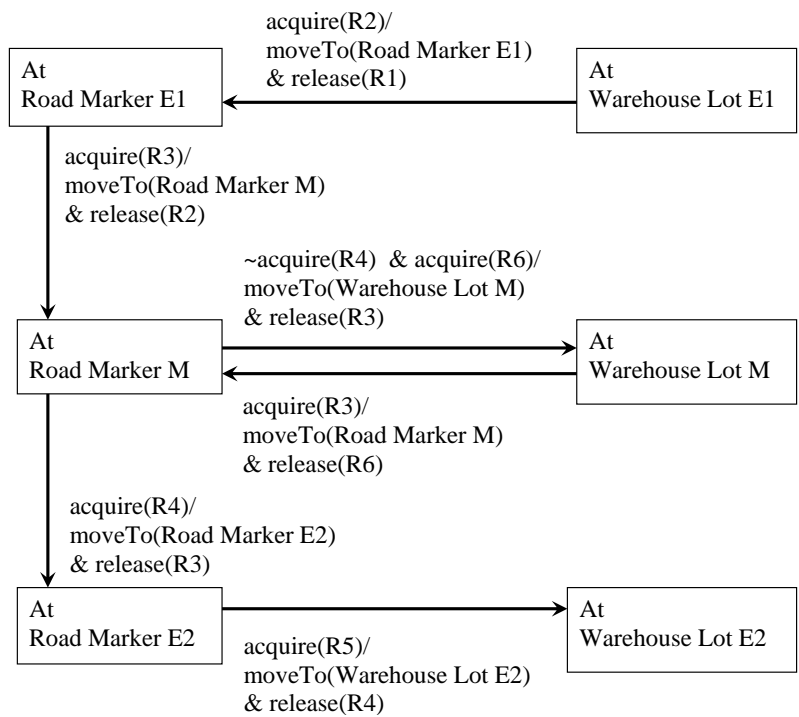

*Fig. 3. State Diagram modified to avoid the deadlock*

#DEFINE N 2

**server**: markerE(agents AMP[N];servers markerM,lotE),
//Edge Road Marker
**services** {tryM[2],tryL,okM[2],okL,takeM,takeL},
//M - going from RMM, L - going from PLE,
//try - test ok access, ok - accept, take - enter
**states** {free,resM,resL,occ},
//free - free, res - reserved, occ - occupied
**actions** {
<i=1..N>            {AMP[i].markerE.tryL, markerE.free}     -> {AMP[i].lotE.ok, markerE.resL},
<i=1..N>            {AMP[i].markerE.takeL, markerE.resL}    -> {AMP[i].markerM.tryE[i], markerE.occ},
<i=1..N>  <j=1..2>  {AMP[i].markerE.okM[j], markerE.occ}    -> {AMP[i].markerM.takeE[j], markerE.free},

<i=1..N>  <j=1..2>  {AMP[i].markerE.tryM[j], markerE.free}  -> {AMP[i].markerM.okE[j], markerE.resM},
<i=1..N>  <j=1..2>  {AMP[i].markerE.tryM[j], markerE.resL}  -> {AMP[i].markerM.notE[j], markerE.resM},
<i=1..N>  <j=1..2>  {AMP[i].markerE.tryM[j], markerE.occ}   -> {AMP[i].markerM.notE[j], markerE.occ},
<i=1..N>            {AMP[i].markerE.takeM, markerE.resM}    -> {AMP[i].lotE.try, markerE.occ},
<i=1..N>            {AMP[i].markerE.okL, markerE.occ}       -> {AMP[i].lotE.take, markerE.free},
}

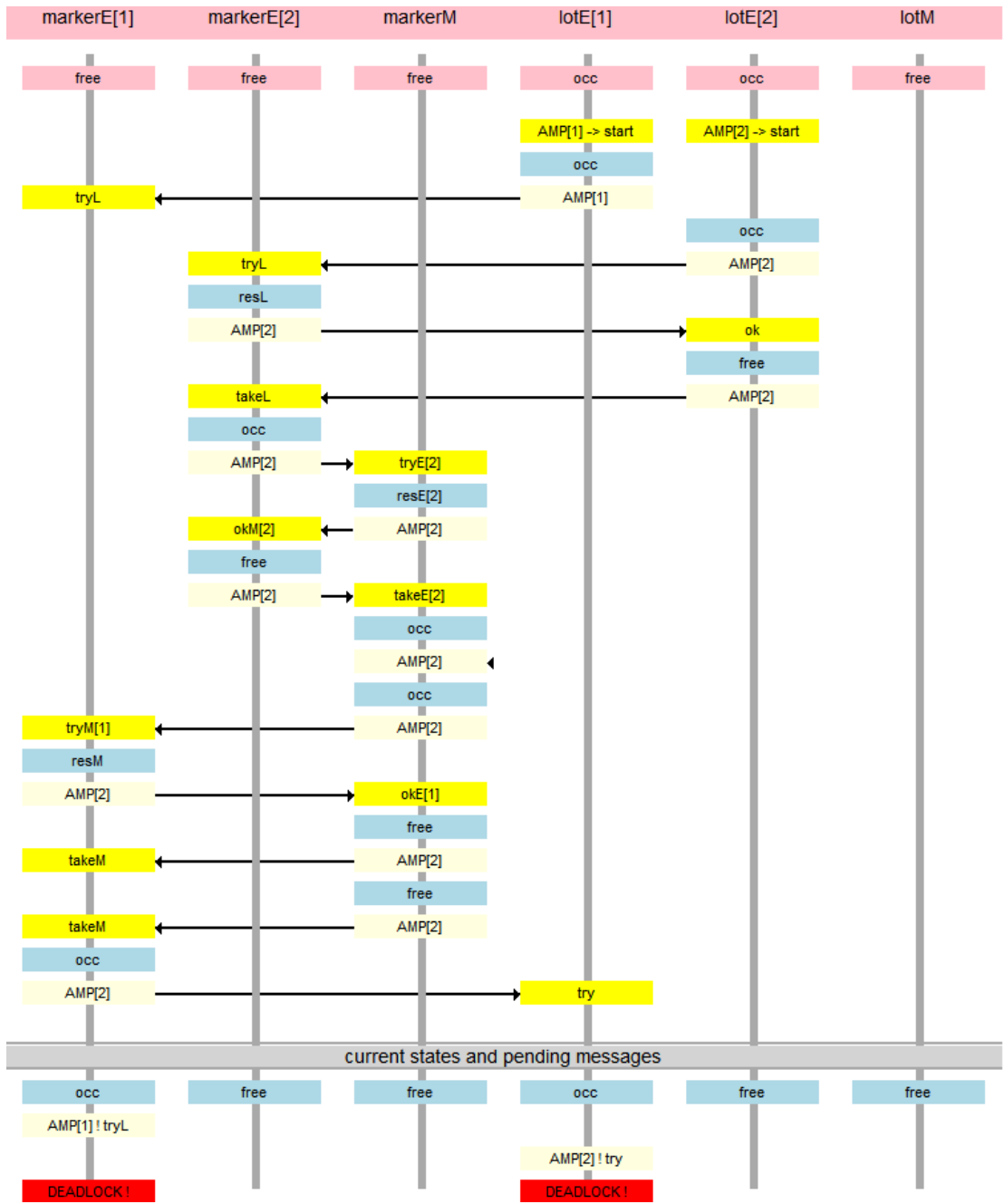

*Fig. 4 Communication structure in a trace of AMPs behavior, leading to the deadlock*

**server**: markerM(agents AMP[N];servers markerE[2],lotM),
//Middle Road Marker
**services**
{tryE[2],tryL[2],okE[2],notE[2],okL[2],takeE[2],takeL[2],switch[2]},
**states** {free,resE[2],resL[2],occ},
**actions** {
//going to ME1 or ME2
<i=1..N>    <j=1..2>    {AMP[i].markerM.tryE[j], markerM.free} -> {AMP[i].markerE[j].okM[j], markerM.resE[j]},
<i=1..N>    <j=1..2>    {AMP[i].markerM.takeE[j], markerM.resE[j]} -> {AMP[i].markerM.switch[3-j], markerM.occ},
<i=1..N>    <j=1..2>    {AMP[i].markerM.switch[j], markerM.occ} -> {AMP[i].markerE[j].tryM[j], markerM.occ},
<i=1..N>    <j=1..2>    {AMP[i].markerM.okE[j], markerM.occ} -> {AMP[i].markerE[j].takeM, markerM.free},

//on a way to ME1 or ME2 may go to LE if MEi occupied
<i=1..N>    <j=1..2>    {AMP[i].markerM.notE[j], markerM.occ} -> {AMP[i].lotM.try[j], markerM.occ},
<i=1..N>    <j=1..2>    {AMP[i].marker2.okL[j], markerM.occ} -> {AMP[i].lotM.take[j], markerM.free},

//going from PL2 - goes to RM1(markerE[1]) or RM3(markerE[2])
<i=1..N>    <j=1..2>    {AMP[i].markerM.tryL[j], markerM.free} -> {AMP[i].lotM.ok[j], markerM.resL[j]},
<i=1..N>    <j=1..2>    {AMP[i].markerM.takeL[j], markerM.resL[j]} -> {AMP[i].markerE[j].tryM[j], markerM.occ},
<i=1..N>    <j=1..2>    {AMP[i].markerM.okE[j], markerM.occ} -> {AMP[i].markerE[j].takeM, markerM.free},
}

**server**:    lotE(agents AMP[N];servers markerE),
//Edge Warehouse Lot
**services** {start,try,ok,take},
**states**   {free,res,occ},

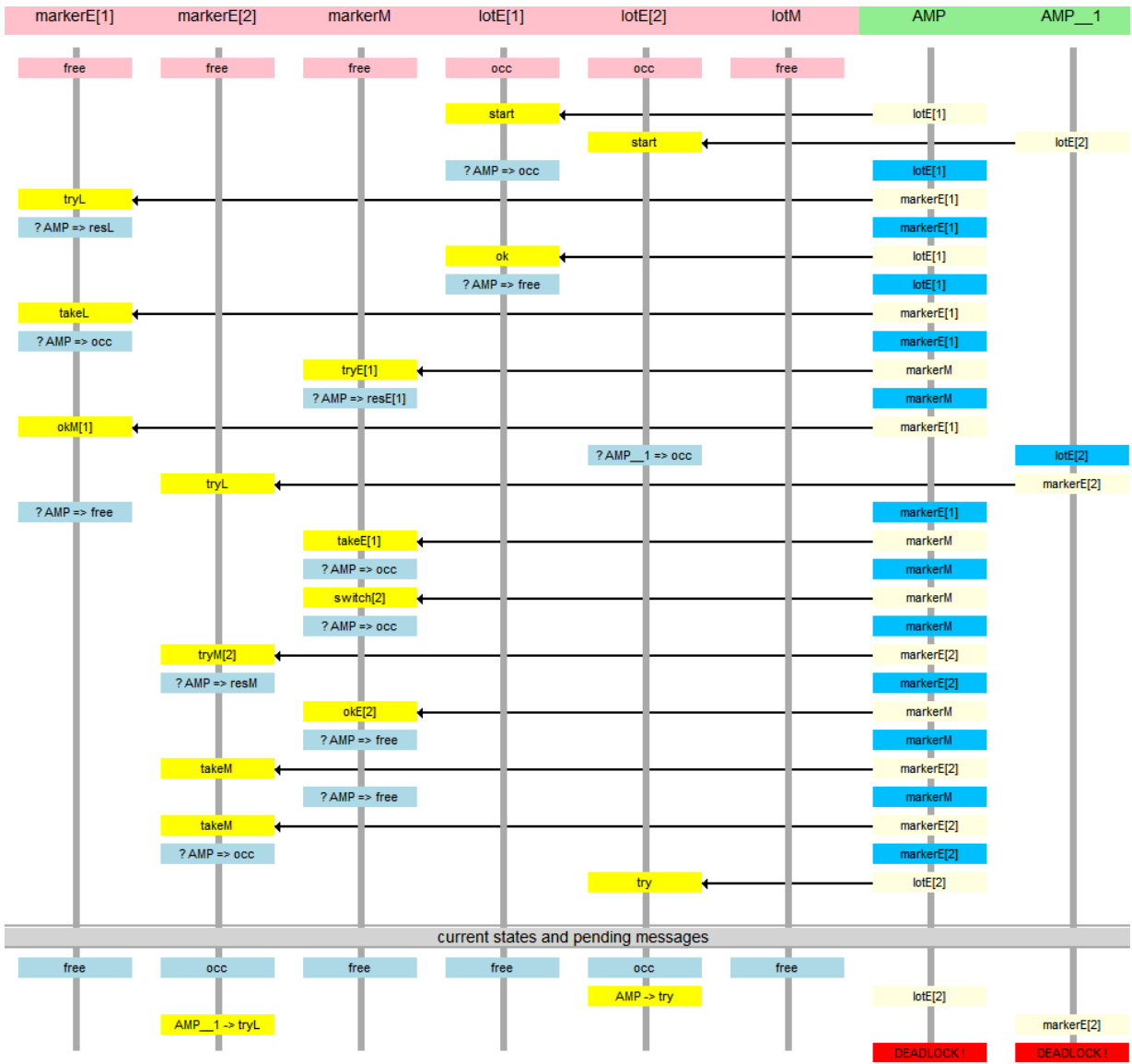

**Fig. 5** *Sequence diagram of AMP(AMP1) moving from Warehouse Lot E1 to Warehouse Lot E2 and AMP__1(AMP2) moving reverse way, leading to the deadlock*

```
actions {
<i=1..N>    {AMP[i].lotE.try, lotE.free} -> {AMP[i].markerE.okL, lotE.res},
<i=1..N>    {AMP[i].lotE.take, lotE.res} -> {lotE.occ},

<i=1..N>    {AMP[i].lotE.start, lotE.occ} -> {AMP[i].markerE.tryL, lotE.occ},
<i=1..N>    {AMP[i].lotE.ok, lotE.occ} -> {AMP[i].markerE.takeL, lotE.free},
}

server:    lotM(agents AMP[N];servers markerM),
//Middle Warehouse Lot
services {try[2],ok[2],take[2]},
states    {free,res[2],occ[2]},
actions {
<i=1..N>    <j=1..2>    {AMP[i].lotM.try[j], lotM.free} -> {AMP[i].markerM.okL[j], lotM.res[j]},
<i=1..N>    <j=1..2>    {AMP[i].lotM.take[j], lotM.res[j]} -> {AMP[i].markerM.tryL[j], lotM.occ[j]},
<i=1..N>    <j=1..2>    {AMP[i].lotM.ok[j], lotM.occ[j]} -> {AMP[i].markerM.takeL[j], lotM.free},
}

servers   markerE[2],markerM,lotE[2],lotM;
agents    AMP[N];

init -> {
<j=1..2>    markerE[j](AMP[1..N],markerM,lotE[j]).free,
            markerM(AMP[1..N],markerE[1,2],lotM).free,
<j=1..2>    lotE[j](AMP[1..N],markerE[j]).occ,
            lotM(AMP[1..N],markerM).free,
<j=1..2>    AMP[j].lotE[j].start,
}.
```

Similar arrangements can be done in a case when an AMP occupies E1 and the other AMP tries to drive from Warehouse Lot E1 to E1. It this case a deadlock occurs, presented in the server view in Fig. 4 and in the agent view in Fig. 5. Both figures are the output from Dedan. Note that in the server view, only two servers out of six are in deadlock. In the agent view, the agents AMP[1] and AMP[2] are renamed to AMP and AMP__1 during automatic conversion from the server view.

## CONCLUSIONS

In this paper, we extended previous studies of cooperating autonomous vehicles to include situations when environmental changes, failures, collisions or even terrorist attacks can cause a malfunctions of the delivery systems. We have presented a novel approach using two-level state diagrams. The lower level diagrams describe computer vision techniques for environmental trigger specifications, movement actions and synchronization primitives. The upper level diagrams allowed us to model advanced interactions between autonomous AMPs. We addressed the problem of ensuring the correctness of AMP interactions. The techniques were presented to verify and analyze combined AMPs' behaviors.

The Dedan verification environment is using model checking techniques, for finding communication deadlocks and resource deadlocks, partial and total. Moreover, the system may be automatically converted from the server view to the agent view, the state space of the system may be observed and simulated, and the system may be converted to Promela (Spin verifier input form, [18]). The described system is still under development. In the near future, a graphical form of verified system representation is planned. A new concept of distributed automata is under development. More advanced forms of verification will be available, using timed automata ([26][27], to verify real-time dependencies), and probabilistic model checking [28]. One of the most advanced feature will be automatic or semi-automatic behavior modification that will significantly improve the dynamic resilience of cooperating autonomous moving platforms.

## ACKNOWLEDGMENT

This research was supported by the Belk Foundation in the case of Bogdan Czejdo and Sambit Bhattacharya.

Authors:

**Bogdan Czejdo**, PhD - Department of Mathematics and Computer Science, Fayetteville State University, Fayetteville, NC 28301, USA, bczejdo@uncfsu.edu

**Sambit Bhattacharya**, PhD - Department of Mathematics and Computer Science, Fayetteville State University, Fayetteville, NC 28301, USA, sbhattac@uncfsu.edu

**Mikołaj Baszun**, PhD - Warsaw University of Technology, Department of Electronics and Information Technology, mbaszun@elka.pw.edu.pl

**Wiktor B. Daszczuk**, PhD – Warsaw University of Technology, Department of Electronics and Information Technology, wbd@ii.pw.edu.pl